\documentclass[a4paper,12pt]{article}
\usepackage{epsfig}
\newcommand{\newc}{\newcommand}
\newc{\lam}{\lambda}
\newc{\ie}{{\it i.e.}}
\newc{\rpv}{\not\!\! R_p}
\newc{\lsim}{\stackrel{<}{\sim}}
\begin{document}
\setlength{\baselineskip}{.7cm}
\title{\textbf{Bounds on Broken $R$-Parity\\ from NOMAD and CHORUS}}
\date{}
\author{ \\Herbi Dreiner$^1$\footnote{dreiner@th.physik.uni-bonn.de}, 
$~$ Giacomo Polesello$^2$\footnote{ Giacomo.Polesello@pavia.infn.it}, 
$~$ Marc Thormeier$^3$\footnote{ thor@thphys.ox.ac.uk  }\\
\\
\\
$^1$ \emph{Physikalisches Institut der Universit\"at Bonn,}\\
\emph{Nu\ss allee 12, 53115 Bonn, Germany}\\ \\
$^2$ \emph{INFN, Sezione di Pavia,}\\\emph{Via Bassi 6, 27100 Pavia, Italy}
\\ \\
$^3$ \emph{Department of Theoretical Physics, University of Oxford,}
\\\emph{1 Keble Road, Oxford OX1 3NP, United Kingdom}}
\maketitle
\begin{abstract}
  $~$\\
  We determine new constraints on the products of two $R$-parity
  violating coupling constants from the NOMAD and CHORUS experiments
  on $\nu_\mu\rightarrow\nu_\tau$ oscillations.  We obtain improved
  results from (a) lepton flavour violating (LFV) meson decays, (b)
  LFV deep-inelastic scattering and (c) a combination of the two.
\end{abstract}
%
%%%%%%%%%%%%%%%%%%%%%%%%%%%%%%%%%%%%%%%%%%%%%%%%%%%%%%%%%%%%%%%%%%%%%%%%%%
%
\section{Introduction}
The superpotential of the $M\!S\!S\!M\!+$$\rpv$ is obtained from the
superpotential of the $ M\! S\! S\!  M $ by adding the following terms
({\it c.f.}  ref.\cite{w})
\begin{eqnarray}\label{superpotential}
\Delta\mathcal{W}_{\not R_p}&=&\frac{1}{2}\varepsilon^{ab}\lam_{ijk}~
{L^i}_a{L^j}_b{E^k}^C+\varepsilon^{ab}\delta^{xy}
{\lam^{\prime}}_{ijk}~{L^i}_{a}{Q^j}_{bx}{{D^{k}}_y}^C\nonumber\\
&&+\frac{1}{2}~~\varepsilon^{xyz}{\lam^{\prime\prime}}_{ijk}~
{{U^{i}}_x}^C{{D^{{j}}}_y}^C{{D^{k}}_z}^C+\varepsilon^{ab}\kappa_i~
{L^i}_a{H^{\mathcal{U}}}_b.
\label{eq:superpot}
\end{eqnarray}
$H,Q,L$ represent the left chiral $SU(2)_W$-doublet superfields of the
Higgses, the quarks and leptons; $U,D,E$ represent the right chiral
superfields of the $u$-type quarks, $d$-type quarks and electron-type
leptons $\ell$, respectively; a superscript $^C$ denotes charge
conjugation; $a,b$ and $x,y,z$ are $SU\!(2)_W$ and $SU\!(3)_C$
indices; $~i,j,k$ and later also $f,l,m$ are generation indices
(summation over repeated indices is implied); $\delta^{xy}$ is the
Kronecker symbol, $\varepsilon^{...}$ symbolizes any tensor that is
totally antisymmetric with respect to the exchange of any two indices,
with $\varepsilon^{12...}=1$. The coupling constants
$\lam_{ijk},{\lam^{\prime\prime}}_{ijk}$ are antisymmetric with
respect to the exchange of the first two/last two indices. The last
term in eq.(\ref{superpotential}) can be rotated away utilizing a
unitary field-redefinition \cite{Hall:1983id}.

Good agreement between the $S\!M$ theory and experiment gives
stringent upper bounds on the extra 45 coupling constants $\lam_{ijk},
{\lam^{\prime}}_{ijk}$ and ${\lam^{\prime \prime}}_{ijk}$, as well as
on products thereof. For a list of references and the processes dealt
with, see e.g. \cite{d,Bhattacharyya:1996nj,add}. The first systematic
analysis of $\not\!\!\!R_p$ bounds was presented in \cite{bgh}, among
other processes also utilizing neutrino deep-inelastic scattering.  We
look at this in considerably more detail here including also lepton
flavour violating meson decays and including explicitly experimental
effects.  We also go beyond \cite{bgh} to allow products of couplings.

Both processes can be tested by neutrino oscillation experiments. In
this article we concentrate on the NOMAD \cite{NOMAD} and CHORUS
\cite{CHORUS} experiments at CERN, for several reasons.  These two
experiments looked for muon-neutrinos oscillating to tau-neutrinos,
thus addressing the products of couplings for second and third
generation leptons, which are less constrained than the first
generation.  The limits obtained by these experiments are the most
stringent to date, thanks to the high statistics running. Lastly,
although the two experiments are dismantled, the two collaborations
are still active, so the numerical factors (see below) which are
necessary to extract firm limits and which are not available in the
published literature can still be evaluated.

\section{Lepton Flavour Violation at CHORUS \& NOMAD}
CERN's SPS fires a proton beam on a Beryllium target.  From the
resulting jet the secondary mesons (mostly pions and kaons) are charge
selected and left to decay into mainly muons and neutrinos.  After the
neutrinos have propagated some distance they are indirectly detected
via neutrino deep-inelastic scattering ($\nu$-DIS) on a nucleon in the
target, producing a charged lepton plus other debris.  The dominating
$S\!M$ process is
\begin{eqnarray}
p+Be~\Longrightarrow~\big(\pi^+~\mbox{or}~K^+\big)~\stackrel{S\!M}{
\longrightarrow}~\mu^C+\nu_\mu& &
~~~~~~~~~\nonumber\\
\nu_\mu&\stackrel{S\!M}\longrightarrow&\nu_\mu\nonumber\\
& &\nu_\mu+N~\stackrel{S\!M}{\longrightarrow}~\mu^-+X,\nonumber
\end{eqnarray}
resulting in an isolated muon and a jet of many hadrons ($X$).  We are
interested in detecting deviations from the $S\!M$ via lepton flavour
violating interactions, in particular those resulting in a final state
tau.  Allowing for at most one non-$S\!M$ process, we have the following
possibilities\footnote{ NOMAD and CHORUS are also sensitive to
  $R$-parity violation via late decaying neutralinos
  \cite{Dedes:2001zi}.}
\begin{eqnarray}
(I)~~~~~~~~~p+Be~\Longrightarrow~\big(\pi^+~\mbox{or}~K^+\big)~\stackrel{
\not R_p}{
\longrightarrow}~\mu^C+\nu_\tau& &
~~~~~~~~~\nonumber\\\nu_\tau&\stackrel{S\!M}{\longrightarrow}&\nu_\tau\nonumber\\
& &\nu_\tau+N~\stackrel{S\!M}{\longrightarrow}~\tau +X,\nonumber\\\nonumber\\
(I\!I)~~~~~~~~~p+Be~\Longrightarrow~\big(\pi^+~\mbox{or}~K^+\big)~\stackrel{S\!M}{
\longrightarrow}~\mu^C+\nu_\mu& &
~~~~~~~~~\nonumber\\\nu_\mu&\stackrel{oscil.}{\longrightarrow}&\nu_\tau
\nonumber\\
& &\nu_\tau+N~\stackrel{S\!M}{\longrightarrow}~\tau +X,\nonumber\\
\nonumber\\
(I\!I\!I)~~~~~~~~~p+Be~\Longrightarrow~\big(\pi^+~\mbox{or}~K^+\big)~\stackrel{S\!M}{
\longrightarrow}~\mu^C+\nu_\mu& &
~~~~~~~~~\nonumber\\\nu_\mu&\stackrel{S\!M}{\longrightarrow}&\nu_\mu\nonumber\\
& &\nu_\mu+N~\stackrel{\not R_p}{\longrightarrow}~\tau+X.\nonumber
\end{eqnarray}
Here we focus on the cases $(I)$ and $(I\hspace{-0.05cm}I
\hspace{-0.05cm}I)$, as well as the combination of $(I)$ and
$(I\hspace{-0.05cm}I \hspace{-0.05cm}I)$.  We do not further discuss
$(I\!I)$ since the case of neutrino oscillations has been treated
extensively, see e.g. \cite{Pesen:zx,p2}.  A rough estimate of the
case $(III)$ was made in \cite{Gninenko:2001id}, which mainly focuses
on other issues.  The detector efficiencies, the spectrum of the
neutrino beam and the convolution of the partonic cross section with
the parton distribution functions (pdf's) were not taken into account.
In addition, the coupling considered in \cite{Gninenko:2001id} also
contributes to $(I)$ and the combined effect must be taken into
account, as we do in section \ref{meson+dis}.  Our bound is a factor
3.3 better.

We are interested in the theoretical prediction of the ratio of the
number of tau-detections $\mathcal{N}_\tau$ over the number of
muon-detections $\mathcal{N}_\mu\,$, to be compared with
$\eta_{exp}\,$, the experimental upper bound on the number of
tau-detections per muon-detection.  The latter is given by CHORUS and
NOMAD from their neutrino oscillation analysis, of which we make
explicit use.  We assume that the new physics does not significantly
affect the muon detection rate and obtain
\begin{equation}\label{zwo} 
\frac{\mathcal{N}_\tau}{\mathcal{N}_\mu}=\frac{\mathcal{N}_\tau^{\not R_p
\mbox{-}decay}+\mathcal{N}_\tau^{oscil.}+\mathcal{N}_\tau^{\not R_p
\mbox{-}D\!I\!S}}{\mathcal{N}_\mu^{S\!M}+\mathcal{N}_\mu^{\not R_p\mbox{-}decay}
+\mathcal{N}_\mu^{oscil.}+\mathcal{N}_\mu^{\not R_p\mbox{-}D\!I\!S}  }
\approx\frac{\mathcal{N}_\tau^{\not R_p\mbox{-}decay}+\mathcal{N}_\tau^{oscil.}
+\mathcal{N}_\tau^{\not R_p\mbox{-}D\!I\!S}}{\mathcal{N}_\mu^{S\!M}}~\leq
~\eta_{exp}.
\end{equation}
Here ${\cal N}_\ell^Y$ is the number of observed leptons $\ell$ which
are due to the process $Y$.  We shall here only analyze $\tau$'s which
originate either from a $\rpv$ meson decay or from $\rpv$-DIS.

The probability that a $\tau$ is detected due to flavour violating
$\nu$-DIS is given by
\begin{eqnarray}
\!\!\!\!
P_{\sigma}&\equiv&\widetilde{\chi}\cdot\frac{\sum\limits_{N=n,p}{\bf N}_{\!N}
\sum\limits_{i=1}^{3}~\sum\limits_{j=1}^{2}~\Big(\chi_{d^iu^j}~\sigma_{\nu_
\mu+N_{d^i}\rightarrow\tau+X}^{\not R_p,~all~E_\nu}~+~\chi_{{u^j}^C{d^i}^C}~
{\sigma}_{\nu_\mu+N_{{u^j}^C}\rightarrow\tau+X'}^{\not R_p,~all~E_\nu} \Big) 
}{\sum\limits_{N=n,p}~{\bf N}_{\!N}~\sum\limits_{i=1}^{3}~\sum\limits_{j=1}^
{2}~\Big(\sigma_{\nu_\mu+N_{d^i}\rightarrow\mu+X}^{S\!M,~all~E_\nu }~+~{
\sigma}_{\nu_\mu+N_{{u^j}^C}\rightarrow\mu+X'}^{S\!M,\;all\;E_\nu} \Big)}\,.
\label{eq:p-sigma}
\end{eqnarray}
Here $P_\sigma=P_\sigma(\lam^{\prime*}_{3jk}\lam^{\prime}_{2ik},\lam^
{\prime*}_{kji}\lam_{2k3})$.  ${\bf N}_{\!N}$ is the number of the
nucleon-type $N$ in the detector target and $\sigma$ denotes a cross
section which has been integrated over the indicated parton
distribution function {\it and} over the incoming neutrino energy
distribution.  The detector efficiency correction factors,
$\widetilde{\chi}$ and $\chi_{q_{i\!n}q_{o\!u\!t}}$, are different for
each physical process,
\begin{eqnarray}\label{777}
\chi_{q_{i\!n}q_{o\!u\!t}}=\frac{\epsilon^{\not R_p}_{\nu_\mu+N_{q_
{i\!n}}\rightarrow  \tau
+X_{q_{o\!u\!t}}}  }{\epsilon^{S\!M}_{\nu_\tau+N_{q_{i\!n}}\rightarrow
  \tau +X_{q_{o\!u\!t}}} }\,,\qquad 
\widetilde{\chi}=\frac{\sigma^{S\!M~~~all~E_\nu}_{\nu_\mu+N_
{q_{i\!n}}{\rightarrow}\mu +X_{q_{o\!u\!t}}} }{\sigma^{S\!M~~~all~E_\nu}
_{\nu_\tau+N_{q_{i\!n}}{\rightarrow}\tau +X_{q_{o\!u\!t}}}}\,.
\end{eqnarray}
$\epsilon$ denotes the efficiency of detecting a tau-neutrino with the
experimental cuts used in the selection analysis, for each respective
process.  The ratio of efficiencies needs to be determined case by
case for each considered analysis by generating a sample of $\rpv$
interactions and applying the experimental cuts on it.  We work with
the target being isoscalar, which is approximately correct for both
experiments \cite{NOMAD,CHORUS}.  Thus the factors ${\bf N}_{\!N}$ in
eq.(\ref{eq:p-sigma}) can be taken outside the sums and cancel.

In order to analyze the lepton flavour violating meson decays we
define the fractions
\begin{eqnarray}
P_\pi(\lam^{\prime*}_{31k}\lam^{\prime}_{21k},\lam ^{\prime*}_{k11}
\lam_{3k2})&\equiv&\frac{\Gamma^{\not R_p}_{\pi\rightarrow\mu^C{\nu_\tau}}}
{\Gamma^{S\!M}_{\pi\rightarrow\mu^C{\nu_\mu}}+\Gamma^{\not R_p}_{\pi\rightarrow
\mu^C{\nu_\tau}}}
\approx \frac{\Gamma^{\not R_p}_{\pi\rightarrow\mu^C{\nu_\tau}}}{\Gamma^{S\!M}
_{\pi\rightarrow\mu^C{\nu_\mu}}}
,\\
P_K(\lam^{\prime*}_{32k}\lam^{\prime}_{21k},\lam^{\prime*}_{k12}\lam_{3k2}
)&\equiv&\frac{\Gamma^{\not R_p}_{K\rightarrow\mu^C{\nu_\tau}}}{\Gamma^
{S\!M}_{K\rightarrow\mu^C{\nu_\mu}}+\Gamma^{\not R_p}_{K\rightarrow\mu^C
{\nu_\tau}}}\approx \frac{\Gamma^{\not R_p}_{K\rightarrow\mu^C{\nu_\tau}}}
{\Gamma^{S\!M}_{K\rightarrow\mu^C{\nu_\mu}}}\,.\label{tres}
\end{eqnarray}
The denominators are dominated by the $S\!M$ decay.  For each decay
mode we have explicitly indicated the possible relevant $\rpv$
coupling constants.

Furthermore, $F_{\pi}$ is the fraction of tau-neutrino events having a
pion as a parent meson which survive the cuts applied for the
oscillation analysis.  The corresponding fraction originating from
kaon decays is denoted $F_K$ and to a good approximation
$F_K=1-F_\pi$.  Using eqs.(\ref{eq:p-sigma}-\ref{tres}) with the
relevant $\rpv$ coupling constants, and since the probabilities are
small, we obtain from eq.(\ref{zwo})
\begin{eqnarray}
F_\pi\cdot P_\pi(\lam^{\prime*}_{31k}\lam^{\prime}_{21k},\lam
^{\prime*}_{k11}\lam_{3k2})~+~\Big(1-F_\pi\Big)\cdot P_K(\lam
^{\prime*}_{32k}\lam^{\prime}_{21k},\lam^{\prime*}_{k12}\lam_{3k2})
~+~~~~&~&\nonumber\\
P_{oscil.}~+~P_\sigma(\lam^{\prime*}_{3jk}\lam^{\prime}_{2ik},
\lam^{\prime*}_{kji}\lam_{2k3})&\leq&\eta_{exp}\,,
\end{eqnarray}
with $i,k=1,2,3$ and $j=1,2$. We assume that the three processes do
not mutually cancel each other, and we can thus determine separate
bounds.  Furthermore, we assume in turn that only two $\rpv$ coupling
constants are non-zero. We thus determine bounds on either the
$\not\!\!R_p$ coupling constants of the meson decay (section
\ref{meson-decay}),
\begin{eqnarray}\label{md}
F_\pi\cdot P_\pi(\lam^{\prime*}_{k11}\lam_{3k2}),\quad
\Big(1-F_\pi\Big)\cdot P_K(\lam^{\prime*}_{k12}\lam_{3k2})
\leq\eta_{exp}\,,
\end{eqnarray}
or the $\not\!\!R_p$ coupling constants of the $\nu$-DIS (section
\ref{nu-dis}),
\begin{eqnarray}\label{psig}
P_\sigma(\lam^{\prime*}_{kji}\lam_{2k3}),~~P_\sigma
(\lam^{\prime*}_{31k}\lam^\prime_{22k}),~~P_\sigma
(\lam^{\prime*}_{31k}\lam^\prime_{23k}),~~P_\sigma
(\lam^{\prime*}_{32k}\lam^\prime_{22k}),~~P_\sigma
(\lam^{\prime*}_{32k}\lam^\prime_{23k})\leq\eta_{exp}\,.
\end{eqnarray}
There are two special cases where a combination of two $\rpv$ coupling
constants contribute to both lepton flavour violating meson decays and
$\nu$-DIS.  Since this is experimentally not distinguishable we
then consider the combined effects, which also lead to an enhanced
sensitivity (section \ref{meson+dis})
\begin{eqnarray}\label{both}
F_\pi\cdot P_\pi(\lam^{\prime*}_{31k}\lam^\prime_{21k})+
P_\sigma(\lam^{\prime*}_{31k}\lam^\prime_{21k}),~~
\Big(1-F_\pi\Big)\cdot P_K(\lam^{\prime*}_{32k}\lam
^\prime_{21k})+P_\sigma(\lam^{\prime*}_{32k}\lam^\prime_{21k}
)\leq\eta_{exp}\,.
\end{eqnarray}
We do not discuss bounds on neutrino oscillation.

We first continue with a discussion of the NOMAD and CHORUS
experiments.  We then discuss the case of an anomalous meson decay in
section \ref{meson-decay}, the $\nu$-DIS is treated in section
\ref{nu-dis}, and the case with meson decay and $\nu$-DIS combined in
section \ref{meson+dis}.  In section \ref{summary} we give a brief
summary.  Note that throughout this article we present the bounds on
the $\not\!\!R_p$ coupling constants in the mass eigenstate base, in
order to avoid model dependent results, see ref.\cite{ag,dm}.  For
theoretical models which predict a hierarchy of $\rpv$ coupling
constants, which we implicitly make use of, see for example
\cite{models,cd}.

%
%%%%%%%%%%%%%%%%%%%%%%%%%%%%%%%%%%%%%%%%%%%%%%%%%%%%%%%%%%%%%%%%%%%%%%%%%%%%%%
%
\section{NOMAD and CHORUS}
\label{exp}
The NOMAD and CHORUS experiments were designed to search for the
oscillation of a muon-neutrino into a tau-neutrino through the
detection of the charged current interaction 
\begin{equation}
\nu_\tau+N\rightarrow \tau+X, 
\end{equation}
followed by the decay of the tau in the detector.  The search was
performed in the muon-neutrino beam of the CERN SPS, which provides an
intense $\nu_\mu$ beam with an average neutrino energy of 27~GeV.
With a fiducial mass in the ton range (770 kg for CHORUS, 2.7 Tons for
NOMAD), and a number of protons on target of $\sim 5\times 10^{19}$,
over a data-taking period of four years, each experiment has collected
on the order of $10^6$ charged current muon-neutrino events ($\nu_\mu^
{CC}$). Once analysis efficiencies have been taken into account, this
yields a sensitivity to the presence of tau-neutrinos in the
muon-neutrino beam at the level of about one part in $10^{4}$ per
experiment.  The two experiments have taken complementary strategies
to the detection of tau decay products.

The \textbf{\emph{N}}$\!\mbox{e}$utrino
\textbf{\emph{O}}\hspace{-0.04cm}scillation
\textbf{\emph{M$\!$a}}$\mbox{g}$netic
\textbf{\emph{D}}$\mbox{e}$tector (NOMAD), described in detail in
ref.\cite{NOMAD}, is based on an active target consisting of drift
chambers immersed in a $0.4\,$T magnetic field. The target is followed
by a set of specialized detectors designed to achieve an efficient
identification of electrons and muons.  This design allows to fully
reconstruct the kinematics of the events, which is used to
discriminate between tau decays occurring in the detector, and
misidentified $\nu_\mu^{CC}$ and $\nu_e^{CC}$ interactions. A set of
detailed analyzes which fully exploit the discrimination power of the
event kinematics separately for each decay channel of the tau have
enabled them to put a stringent limit on the muon-neutrino to
tau-neutrino oscillation probability. The final NOMAD result at 90\%
CL is \cite{as}
\begin{equation}
P_{oscil.}^{N\!O\!M\!A\!D}<1.63 \times 10^{-4}.
\end{equation}

The \textbf{\emph{C}}$\!\mbox{E}$RN \textbf{\emph{H}}$\!\mbox{y}$brid
\textbf{\emph{O}}$\mbox{s}$cillation \textbf{\emph{R}}$\mbox{e}$search
Apparat$\!$\textbf{\emph{us}} (CHORUS), described in detail in
ref.\cite{CHORUS}, uses a hybrid setup, with a target based on nuclear
emulsions, followed by a spectrometer system and  a calorimeter.
This allows the measurement of the momentum of the particles emerging
from the target region, and the identification of the muons.

The information of the electronic detectors is used to apply loose
kinematic selections, which define the data sets on which to perform
the tau search in the emulsion stack. The decay modes of the tau into
a muon or a single hadron are identified by the detection of a {\em
  kink}, i.e. a track from the interaction vertex showing a change in
direction after a short path, of the order of $\gamma c\tau_\tau={\cal
  O}(1\,mm)$.  The latest CHORUS results, based on the analysis of a
fraction of the events, see ref.\cite{es}, give no candidate observed,
yielding a limit:
\begin{equation}
P_{oscil.}^{C\!H\!O\!R\!U\!S}<3.4 \times 10^{-4}.
\end{equation}
It should be noted that the CHORUS oscillation limit is obtained
with a different statistical treatment than the NOMAD one. If the same
treatment were applied, the CHORUS result would be roughly
$P_{oscil.}^{C\!H\!O\!R\!U\!S}<2.1 \times 10^{-4}$, see
ref.\cite{roberto}.

No official combined limit from the two experiments is available.  An
exercise of combining the results, based on the statistical techniques
used in NOMAD \cite{roberto} , yields
\begin{equation}\label{combi}
P_{oscil.}^{combined}<5 \times 10^{-5}.
\end{equation}
We thus use $\eta_{exp}=5 \times 10^{-5}$.

The two experiments have also obtained limits on the
electron-neutrinos oscillating to tau-neutrinos, based on the $\sim
1\%$ contamination of electron-neutrinos, in the muon-neutrino beam.
The limits are about two orders of magnitude worse than the previously
discussed case, which is why we do not treat electron-neutrinos in
this paper.  The extension is however straightforward.
%
%%%%%%%%%%%%%%%%%%%%%%%%%%%%%%%%%%%%%%%%%%%%%%%%%%%%%%%%%%%%%%%%%%%%%%%%%%%%%%
%
\section{$\big(\pi^+~\mbox{or}~K^+\big)\stackrel{\not R_p}{
\longrightarrow}\mu^C+\nu_\tau$}
\label{meson-decay}
Starting from eq.(\ref{md}), the results in section 2.1  of  
ref.{\cite{dpt}} lead to 
\begin{eqnarray}
\frac{\Gamma^{\not R_p}_{\pi\rightarrow\mu+\nu_\tau}}{\Gamma^{S\!M}
_{\pi\rightarrow\mu+\nu_\mu}}=|K_{1123}|^2\leq
\frac{\eta_{exp}}{F_\pi},
\end{eqnarray}
with 
\begin{equation}
K_{1123}=\frac{-~m^2_{\pi}}{2\sqrt{2}~G_F~|V_{11}|~m_{\mu}(m_{u}+m_{d})}
~\frac{{\lam}^{\prime*}_{k11}~\lam_{3k2}}{m^2_{
\widetilde{{\ell_L}^k}}}\,.
\label{Kpi}
\end{equation}
$G_F$ is the Fermi constant and $V_{ji}$ is an element of the
CKM-matrix, $m$ symbolizes a mass.  Here we have neglected the
correction factors due to higher order electroweak leading logarithms,
short distance $QC\!D$ corrections, and structure dependent effects.
The analogous constant to eq.(\ref{Kpi}) for kaon decay, $K_{2123}$ is
obtained from $K_{1123}$ by replacing $m_\pi\rightarrow m_K$,
$m_d\rightarrow m_s$ and ${\lam}^{\prime*}_{k11}~\lam_{3k2}
\rightarrow \lam^{\prime*}_{k12}\lam_{3k2}$.

\begin{figure}[t]
\centerline{\hbox{\psfig{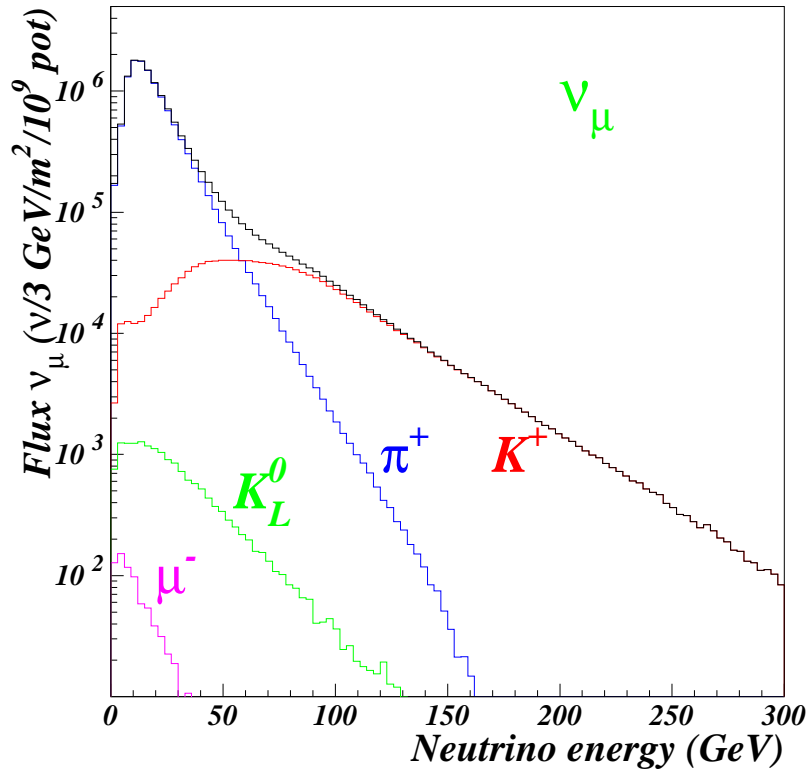}}
}
\caption{{\it Spectrum of the muon-neutrino beam at the NOMAD detector,
    from \cite{spectrum}. We show the muon neutrino flux originating
    from different particle decays. The two components of the neutrino
    spectrum, the one coming from pion and the other one coming from
    kaon decay, are separately shown on the graph.  }}
\label{fig:nuspectra}
\end{figure}

With $G_F=(0.116639 \pm 0.000001)\times(100~\mbox{GeV})^{-2}$,
$m_\mu=(105.6583568\pm5.2\times10^{-6})$ MeV, $m_\pi=(139.57018
\pm0.00035)$ MeV, $m_u+m_d=(8.5\pm3.5)$ MeV, $|V_{11}|=0.9750
\pm0.0008$, $m_{K}=(493.677\pm0.016)$ MeV, $m_s=(122.5\pm47.5)$ MeV,
$m_s=(21\pm4)~m_d$, and $|V_{12}|=0.222\pm0.004$ \cite{pdg}, one gets,
using central values,
\begin{eqnarray}
|\lam^{\prime*}_{k11}\lam_{3k2}|&\leq&0.014\,\sqrt{\frac{\eta_{exp}}
{ F_\pi}}\left(\frac{m_{\widetilde{{\ell_L}^k}}}{100\,\rm{GeV}}
\right)^2,\\
|\lam^{\prime*}_{k12}\lam_{3k2}|&\leq&0.0039\,\sqrt{\frac{\eta_
{exp}}{1- F_\pi}} \left(\frac{m_{\widetilde{{\ell_L}^k}}}{100\,\rm{GeV}}
\right)^2\!\!.
\end{eqnarray}
The value of $F_\pi$ cannot be evaluated with the published numbers.
It is in fact a number which depends on the neutrino energy, and
varies between $\sim0$ for the lowest accessible neutrino energies to
$\sim1$ for the highest neutrino energy, as can be seen in
Fig.\ref{fig:nuspectra}. This factor in turn needs to be convoluted
with the efficiency of the analysis cuts at a given energy, which is
also a function of the specific analysis.

The dependence on the neutrino energy of the global analysis
efficiency for CHORUS is given in \cite{es}.  By convoluting the
analysis efficiency with the different components of the neutrino
spectra shown in Fig.~\ref{fig:nuspectra}, the approximate value for
$F_\pi$ obtained in this way is $\sim0.92$.  We expect that the
effective $F_\pi$ from the combination of the different analyses from
NOMAD to be lower than this value.  In fact CHORUS, for an equivalent
limit on the oscillation probability, is able to exclude a
significantly lower value of $\Delta m^2_\nu$, the difference of the
mass squared of the two neutrino flavours \cite{as,es}.  This means
\cite{pdg} that the average energy of the neutrinos surviving the
analysis cuts is higher for NOMAD than for CHORUS.  As can be seen
from Fig.~\ref{fig:nuspectra}, a harder energy spectrum enhances the
fraction of events for which the parent meson is a $K$.  We extract a
conservative limit from the available information by assuming that the
NOMAD and CHORUS results contribute with a similar weight to the
combined limit.  For NOMAD we take $F_\pi$ varying between $\sim$0.1
and $\sim$0.9. The final range for the effective value of $F_\pi$ in
our combined analysis is therefore:
\begin{equation}
0.5\leq F_\pi \leq 0.9.
\end{equation}
To be as conservative as possible we use the lower bound when dealing with
the pion and the upper bound when dealing with the kaons. With eq.(\ref{combi}) we 
therefore obtain
\begin{eqnarray}
|\lam^{\prime*}_{211}\lam_{322}|\leq1.4\times10^{-4}
~\Bigg(\frac{m_{\widetilde{{\ell_L}^k}}}{100~\mbox{GeV}}\Bigg)^2,  \nonumber\\
|\lam^{\prime*}_{k12}\lam_{3k2}|\leq 8.8\times10^{-5}
~\Bigg(\frac{m_{\widetilde{{\ell_L}^k}}}{100~\mbox{GeV}}\Bigg)^2.
\end{eqnarray}
Using the product of the bounds on single coupling constants
\cite{add}, one gets a stricter bound on $|\lam^{\prime*}_{111}
\lam_{312}|$ than we would obtain here.

%
%%%%%%%%%%%%%%%%%%%%%%%%%%%%%%%%%%%%%%%%%%%%%%%%%%%%%%%%%%%%%%%%%%%%%%%%%%%%%%
%
\section{$\nu_\mu+N\stackrel{\not R_p}{\longrightarrow}\tau+X$}
\label{nu-dis}
In order to determine a bound from $\nu$-DIS we must compute the cross
section
\begin{equation}\label{ui}
{\sigma}_{\nu_\mu+N_{q_{i\!n}}\rightarrow\ell^f+X_{q_{o\!u\!t}}}^
{all~E_\nu}= \int\limits_
{\!\!E_{\nu_{\!\mu} m\!i\!n}^{l\!a\!b}}^{~~E_{\nu_{\!\mu} m\!a\!x}^
{l\!a\!b}}  {\sigma}_{\nu_\mu+N_{q_{i\!n}}\rightarrow\ell^f+X_{q_{o\!u\!t}}}\!
(s(E_{\nu_\mu}^{lab}))  ~~g(E_{\nu_\mu}^{lab})~~dE_{\nu_\mu}^{lab}\,,
\end{equation}
for both the $S\!M$ process ($f=2$) and the $\rpv$\ process ($f=3$). $
g(E_{\nu_\mu}^{lab})$ is the incoming neutrino energy spectrum (in the
lab frame) and is given in ref.\cite{spectrum}.  $s\equiv(p_{\nu_\mu}
+p_N)^2=2E_{\nu_\mu} ^{lab}m_N+m_N^2$ is the center-of-mass energy.
We perform the above integral numerically.  The cross section in
eq.(\ref{ui}) implicitly contains an integral over the quark
distribution functions, which we discuss below.  Explicitly in terms
of the cross sections we have from eq.(\ref{psig})
\begin{eqnarray}\label{fr}
\frac{\widetilde{\chi}~\sum\limits_{N,i,j}\int\limits_
{\!\!E_{\nu_{\!\mu} m\!i\!n}^{l\!a\!b}}^{~~E_{\nu_{\!\mu} m\!a\!x}^
{l\!a\!b}}  \!\!\!\!\!\!g(E_{\nu_\mu}^{lab})\Big(\chi_{d^iu^j}~\sigma^{\not R_p}
_{\nu_\mu+N_{d^i}{\rightarrow}\tau+X_{u^j}}\![s]+\chi_{{u^j}^C{d^i}^C}~{\sigma}^
{\not R_p}_{\nu_\mu+N_{{u^j}^C}{\rightarrow}\tau+X_{{d^i}^C}}\![s] \Big)
~dE_{\nu_\mu}^{lab}   }
{\sum\limits_{N,i,j}\int\limits_
{\!\!E_{\nu_{\!\mu} m\!i\!n}^{l\!a\!b}}^{~~E_{\nu_{\!\mu} m\!a\!x}^
{l\!a\!b}}  \!\!\!\!\!\!g(E_{\nu_\mu}^{lab})\Big(\sigma^{S\!M}_{\nu_\mu+N_{d^i}
{\rightarrow}\mu+X_{u^j}}\![s]+{\sigma}^{S\!M}_{\nu_\mu+N_{{u^j}^C}\rightarrow
\mu+X_{{d^i}^C}}\![s] \Big)
~dE_{\nu_\mu}^{lab}}\leq\eta_{exp}.~~
\label{Psigeta}
\end{eqnarray}
We calculate the cross sections for both the $S\!M$ and the $\rpv$\ 
case by first computing the partonic cross sections $\sigma_{\nu_
  \mu+{q_{i\!n}}\rightarrow\ell^f+{q_{o\!u\!t}}}$.  With $\hat{s}
\equiv (p_{\nu_\mu}+p_{q_{i\!n}})^2,$ and $t\equiv(p_ {\nu_\mu}-
p_{\ell^f})^2$ the averaged matrix elements squared in the $S\!M$ are
given by
\begin{eqnarray}
\langle|\mathcal{M}_{\nu_\mu+d^i\stackrel{S\!M}{\longrightarrow}\mu+u^j}|^2
\rangle&=& 16~G_F^2~
|V_{ji}|^2~(\hat{s}-m_{d^i}^2)~(\hat{s}-m_\mu^2-m_{u^j}^2)\,,\label{mat1}\\
\langle|\mathcal{M}_{\nu_\mu+{u^j}^C\stackrel{S\!M}{\longrightarrow}\mu+{d^i}^C}
|^2\rangle & = 
&16~G_F^2~|V_{ji}|^2~(\hat{s}+t-m_{u^j}^2-m_\mu^2)~(\hat{s}+t-m_{d^i}^2)\,.
\label{mat2}
\end{eqnarray}
In the $\rpv$\ case,  we obtain 
\begin{eqnarray}
\langle|\mathcal{M}_{\nu_\mu+d^i\stackrel{\not R_p}{\longrightarrow}
\tau+u^j}|^2\rangle & = & \frac{1}{2}~ \Bigg|\sum_k~\frac{\lam^{\prime*}_{3jk}~\lam
^\prime_{2ik}}{m_{\widetilde{{d_R}^k}}^2} \Bigg|^2 ~(\hat{s}-m_{d^i}
^2)~(\hat{s}-m_\tau^2-m_{u^j}^2)\nonumber\label{mat3}\\
\hspace{-1cm}&+&\mbox{Re}\Bigg[\Bigg(\sum_k~\frac{\lam
^{\prime*}_{3jk}~\lam^\prime_{2ik}}{m_{\widetilde{{d_R}^k}}
^2}\Bigg) \cdot \Bigg(\sum_k~\frac{\lam^{\prime*}_{kji}~\lam_{2k3}}
{m_{\widetilde{{\ell_L}^k}}^2} \Bigg)^{\!\!*}\Bigg]~m_{d^i}~m_{\tau}~
(\hat{s}+t-m_{\tau}^2-m_{d^i}^2)~\nonumber\\
&+&\frac{1}{2}~\Bigg|\sum_k~\frac{\lam^{\prime*}_{kji}~\lam_{2k3}}
{m_{\widetilde{{\ell_L}^k}}^2} \Bigg|^2~(m_{d^i}^2+m_{u^j}^2-t)~
(m_{\tau}^2-t)\,,
\end{eqnarray}
\vspace{-0.5cm}
\begin{eqnarray}
\langle|\mathcal{M}_{\nu_\mu+{u^j}^C\stackrel{\not R_p}{\longrightarrow}\tau+{d^i}^C}
|^2\rangle & = &
\frac{1}{2}~ \Bigg|\sum_k~\frac{\lam^{\prime*}_{3jk}~\lam^\prime_{2ik}}
{m_{\widetilde{{d_R}^k}}^2} \Bigg|^2 ~(\hat{s}+t-m_\tau^2-m_{u^j}^2)~(\hat{s}
+t-m_{d^i}^2)\nonumber\\
&+&\mbox{Re}\Bigg[\Bigg(\sum_k~\frac{\lam^{\prime*}_{3jk}~\lam^
\prime_{2ik}}{m_{\widetilde{{d_R}^k}}^2}\Bigg)\cdot  \Bigg(\sum_k~\frac{\lam
^{\prime*}_{kji}~\lam_{2k3}}{m_{\widetilde{{\ell_L}^k}}^2} \Bigg)^{\!\!*}\Bigg]    
~m_{d^i}~m_{\tau}~(\hat{s}-m_{u^j}^2)~\nonumber\\
&+&\frac{1}{2}~\Bigg|\sum_k~\frac{\lam^{\prime*}_{kji}~\lam_{2k3}}
{m_{\widetilde{{\ell_L}^k}}^2} \Bigg|^2   ~(m_{d^i}^2+m_{u^j}^2-t)
(m_{\tau}^2-t)\,. \label{mat4}
\end{eqnarray}
$\langle...\rangle$ denotes the average/sum over initial/final state
spins.  The differential cross section in the partonic CM-frame is
given by
\begin{equation}
\frac{d\sigma_
  {\nu_\mu+q_{i\!n}\rightarrow\ell^f+q_{o\!u\!t}}}{d\cos\vartheta^{C\!M}
  }=\frac{\langle|\mathcal{M}_{\nu_\mu+q_{i\!n}\rightarrow\ell^f
    +q_{o\!u\!t}}|^2\rangle~|\vec{p}_{\ell^f}|^{C\!M}}{32~\pi~\hat{s}~
  |\vec{p}_{\nu_\mu}|^{C\!M}}\,, \label{diffXsect}
\end{equation}
where $\vartheta^{C\!M}$ denotes the angle between the incoming neutrino
and the outgoing charged lepton in the partonic CM-frame.  In order to
obtain the nuclear cross section we have to fold the partonic cross
section with the pdf's $q_{i\!n}^N(x,t)$
\begin{equation}
\sigma_{\nu_\mu+N_{q_{i\!n}}\rightarrow\ell^f+X_{q_{o\!u\!t}}}(s)=
\int\limits_{\!x_{m\!i\!n}}^{~~x_{m\!a\!x}} \int\limits_{\!1~~}^{~-1}
~\frac{d\sigma_{\nu_\mu+q_{i\!n}\rightarrow\ell^f+q_{o\!u\!t}}}{d
\cos\vartheta^{C\!M}}(\hat{s},t)~q_{i\!n}^N(x,t)~d\cos\vartheta^{C\!M}~dx\,.
\label{Xsect}
\end{equation}  
We use the pdf's given in ref.\cite{mrst}. The Mandelstam variable $t$
is given by\footnote{For the sake of generality we shall at the moment
  not work with $m_{q_{i\!n}}=0$, although it is required by the
  parton model.}
\begin{equation}
t=\frac{1}{2\hat{s}}\Big((m_{\ell^f}^2+m_{q_{o\!u\!t}}^2-\hat{s})\hat{s}
+(m_{\ell^f}^2-m_{q_{o\!u\!t}}^2+\hat{s})m_{q_{i\!n}}^2+(\hat{s}-
m_{q_{i\!n}}^2) \cos \vartheta^{C\!M} \sqrt{...\phantom{|}}~\Big)\,.
\label{t}
\end{equation}
and $\sqrt{...\phantom{|}}\equiv\sqrt{m^4_{\ell^f}+(m_{q_{o\!u\!t}}^2
  -\hat{s}) ^2-2~m^2_{\ell^f}(m_{q_{o\!u\!t}}^2+\hat{s})}$.
Furthermore $x$ is the Bjorken scaling variable, defined as $p_{q_
  {i\!n}}=xp_N$. By definition $x_{max}=1$. $x$ is given explicitly in
terms of the momenta by
\begin{equation}
x=\frac{s-
  m_N^2}{2~m_N^2}~\Bigg[\sqrt{1+\Big(\frac{2(p_{\ell^f}+ p_{q_{out }})
    ~m_N}{s-m_N^2}\Big)^2}~-~1\Bigg]\,.
\end{equation}
We obtain the minimum allowed value, $x_{min}$, when the outgoing
fermions are at rest in the partonic CM-frame
\begin{eqnarray}\label{xmiin}
x_{min}=\frac{s-m_N^2}{2~m_N^2}~\left[\sqrt{1+\Bigg(\frac{2(m_{\ell^f}
+m_{q_{o\!u\!t}})~m_N}{s-m_N^2}\Bigg)^2}~-~1\right].
\label{xmin}
\end{eqnarray}
For each matrix element squared in eqs.(\ref{mat1})-(\ref{mat4}), we
can compute the total cross section in eq.(\ref{Xsect}) using the
differential cross section in eq.(\ref{diffXsect}) as well as the
formul{\ae} of eqs.(\ref{t}) through (\ref{xmin}). The $S\!M$ cross
sections as well as the relevant $\rpv$\ cross section are then
inserted in eq.(\ref{Psigeta}) in order to obtain a bound on the
$\rpv$ coupling constants.

As stated above, we assume that in any given case only two $\rpv
$-couplings are non-zero. There are then no interference terms in the
matrix elements squared given in eqs.(\ref{mat3}) and (\ref{mat4}) and
the product of the couplings factor out of all the integrals. We thus
directly obtain a bound on the product of the $\rpv$ coupling
constants considered.\footnote{From now we work with $m_{q_{i\!n}}=0$,
  as required by the parton model.}

We determine explicit bounds with the following numerical values
$m_c=(1250\pm100)$ MeV, $m_b=(4200\pm200)$ MeV, $m_\tau=(1776
.99\pm0.29)$ MeV, $m_p=(938.271998\pm0.000038)$ MeV, $m_n=(939.565
330\pm0.000038)$ MeV, $|V_{13}|=0.0035\pm0.0015$, $|V_{21}|=0.222
\pm0.003 $, $|V_{22}|=0.941 \pm0.008 $, $|V_{23}|=0.040 \pm0.03 $,
\cite{pdg}. We then obtain
\begin{eqnarray}
|\lam^{\prime*}_{k11}~\lam_{2k3}  |&\leq&\frac{1}{\sqrt{0.43~\chi_{du}
+0.036~\chi_{u^Cd^C}}  }~\times~\sqrt{\frac{\eta_{exp}}{{\widetilde{\chi}}}}
~\Bigg(\frac{m_{\widetilde{{\ell_L}^k}}}{100~\mbox{GeV}}\Bigg)^2.
\label{general}
\end{eqnarray}
The coefficients of the $\chi_{q_{i\!n}q_{o\!u\!t}}$ depend on the
pdf's and we determined them numerically.  Thus each process listed in
eq.(\ref{psig}) will have different coefficients.  The other
processes, however, do not lead to improved bounds beyond those
published in the literature \cite{add}, as we discuss below. For
completeness we list the corresponding set of equations in the
appendix.

The factor $\widetilde{\chi}$ is $1/0.48$ for the NOMAD DIS analyses
and $1/0.53$ for CHORUS (see ref.\cite{as} and ref.\cite{es}).  We
shall hence work with
\begin{eqnarray}
\widetilde{\chi}=2.
\end{eqnarray}
The standard model $\nu_\tau^{CC}$ cross section is dominated by 
$\nu_\tau d\rightarrow \tau^- u$. One can therefore make the simplifying
assumption 
\begin{equation}\label{33}
\chi_{du}\sim 1\,.
\end{equation}
There is no reliable estimate for the other quark combinations,
$\chi_{q_{i\!n}q_{o\!u\!t}}$. In fact the difference of the momentum
distribution of the quarks in the nucleon, with respect to the valence
$d$ quark, and/or the presence of heavy ($c$ or $b$) quarks in the
final state significantly alters the kinematic distributions of the
final state products.  It would therefore be necessary to perform a
detailed simulation of the experimental analysis to extract the
efficiency values.  Therefore, we only explicitly calculate the case
(with $\chi_{u^Cd^C}=0$)
\begin{equation}
|\lam^{\prime*}_{k11}~\lam_{2k3}  |\leq 0.0076 \,
\Bigg(\frac{m_{\widetilde{{\ell_L}^k} }}{100~\mbox{GeV}}\Bigg)^2.
\end{equation}
This is to be compared with the product of single coupling bounds
summarized in ref.\cite{add}.  For $k=1$ the product of single bounds
is significantly stricter. $k=2$ is not allowed due to the
anti-symmetry in the $LLE^C$ operators. For $k=3$ the product of
single coupling bounds is virtually identical to the above bound.
However, there the bound on $\lam'_{311}$ depends on the squark mass
$m_{\widetilde{d_R}}$ and the bound on $\lam_{233}$ depends on the
right handed slepton mass.  If $m_{\widetilde {d_R}}\gg
m_{\widetilde{{\tau_L}}} $, our bound will be the best bound.

The next most promising case for an additional new limit is
$\lam^{\prime*}_{k21}\lam_{2k3}$, with an in-going valence $d$-quark
and an out-going charm quark. Although we cannot quantify $\chi_{dc}$,
we can evaluate the limit under optimistic
assumptions.  For
$\chi_{dc}=1$  and $\chi_{c^Cd^C}=0$ we obtain
\begin{equation}
|\lam^{\prime*}_{k21}\lam_{2k3}| \lsim 0.006\,
\Bigg(\frac{m_{\widetilde{{\ell_L}^k} }}{100~\mbox{GeV}}\Bigg)^2.
\end{equation}
This is comparable to the product of the single couplings limits for
$k=1$, which is $0.002$ and somewhat better for $k=3$: 0.035.  The
bounds on the other products are not competitive.  In the appendix we
have summarized the formul{\ae} for the bounds from $\nu$-DIS,
explicitly retaining $\chi_{q_{i\!n}q_{o\!u\!t}},
\eta_{exp},\widetilde{\chi}$.

%
%%%%%%%%%%%%%%%%%%%%%%%%%%%%%%%%%%%%%%%%%%%%%%%%%%%%%%%%%%
%
\section{$\big(\pi^+~\mbox{or}~K^+\big)\stackrel{\not R_p}{
\longrightarrow}\mu^C+\nu_\tau$ combined with $\nu_\mu+N\stackrel{\not R_p}{
\longrightarrow}\tau+X$}
\label{meson+dis}
Here we use eq.(\ref{both}), i.e.  we combine DIS and meson decays.
For the DIS analysis we use eq.(\ref{33}).  For the meson decays we
again make use of the results stated in ref.\cite{dpt} section 2.1.
The conservative approach we follow is to set $F_\pi$ to the smallest
value in the established range for each of the expressions stated
below, and furthermore to set $\chi_{u^Cd^C}=\chi_{dc}=\chi_{c^Cd^C}
=0$.  We obtain the limits 
\begin{eqnarray}
|\lam^{\prime*}_{31k}~\lam^\prime_{21k}| &\leq&\! 
\frac{0.117~\sqrt{32~\eta_{exp}}}
{\sqrt{\frac{F_\pi}{|V_{11}|^2}+\widetilde{\chi}(0.49~\chi_{du}
+0.012~\chi_{u^Cd^C})}  }
~\Bigg(\frac{m_{\widetilde{d_R}^k}}{100~\mbox{GeV}}\Bigg)^2\nonumber\\& &~~~~~~
=0.0038~\Bigg
(\frac{m_{\widetilde{d_R}^k}}{100~\mbox{GeV}}\Bigg)^2\!\!\!, \\ &&\nonumber\\
&&\nonumber\\
|\lam^{\prime*}_{32k}~\lam^\prime_{21k}  |
&\leq&\! \frac{0.117~\sqrt{32~\eta_{exp}}}
{\sqrt{\frac{1-F_\pi}{|V_{12}|^2}+\widetilde{\chi}(0.43~\chi_{dc}
+0.0015~\chi_{c^Cd^C})}  }
\Bigg(\frac{m_{\widetilde{d_R}^k}}{100~\mbox{GeV}}\Bigg)^2\nonumber\\& &~~~~~~
=0.0027 \Bigg(\frac{m_
{\widetilde{d_R}^k}}{100~\mbox{GeV}}\Bigg)^2\!.~~~~~ 
\end{eqnarray}
Note that in the second case we combined
$K^+\rightarrow\mu^C+\nu_\tau$ with $\nu_\mu+d
\rightarrow\tau+c\Big/\nu_\mu+c^C\rightarrow\tau+d^C$. Both bounds are
improvements compared to products of bounds on single coupling
constants: respectively 0.0064 and 0.031 \cite{add}.

%
%%%%%%%%%%%%%%%%%%%%%%%%%%%%%%%%%%%%%%%%%%%%%%%%%%%%%%%%%%%
%
\section{ Summary and Conclusion}
\label{summary}
We have investigated the bounds that can be obtained on the product of
$R$-parity violating coupling constants by employing the neutrino
oscillation analyses of the NOMAD and CHORUS experiments. They are
sensitive to lepton number violating meson decays leading to tau
neutrinos and to neutrino deep inelastic scattering in which tau
leptons are produced in the final state. We obtain the following
bounds which are more sensitive than the product of existing single
coupling bounds, and thus present the best bounds on these coupling
constant combinations
\begin{eqnarray}
|\lam^{\prime*}_{211}~\lam_{322}  |&\leq& 1.4\times10^{-4} 
\Bigg(\frac{m_{\widetilde{\ell_L}
^2}}{100~\mbox{GeV}}\Bigg)^2,\\
|\lam^{\prime*}_{k12}~\lam_{3k2}  |&\leq&8.8\times10^{-5} 
\Bigg(\frac{m_{\widetilde{\ell_L}
^k}}{100~\mbox{GeV}}\Bigg)^2,\\
|\lam^{\prime*}_{31k}~\lam^\prime_{21k}  |&\leq& 
3.8\times10^{-3} \Bigg(\frac{m_{\widetilde{d_R}
^2}}{100~\mbox{GeV}}\Bigg)^2,\\
|\lam^{\prime*}_{32k}~\lam^\prime_{21k}  |&\leq& 
2.7\,\times10^{-3} \Bigg(\frac{m_{\widetilde{d_R}
^k}}{100~\mbox{GeV}}\Bigg)^2.
\end{eqnarray}
For completeness we list the formul{\ae} for all processes
corresponding to eq.(\ref{general}) in the appendix.  Even for
optimistic estimates of the efficiencies, the resulting bounds are
weaker than existing bounds or only a marginal improvement.

%
%%%%%%%%%%%%%%%%%%%%%%%%%%%%%%%%%%%%%%%%%%%%%%%%%%%%%%%%%%%
%
\section{Acknowledgments}
We thank Ulrich Langenfeld for useful discussions on  numerical 
computing and Dick Roberts for valuable advice on MRST2001. 
We thank Roberto Petti for enlightening discussions on the NOMAD
results. M.T. gratefully 
acknowledges the kind hospitality of the Physikalisches Institut at  
Universit\"at Bonn, and the financial support of the  Evangelisches 
Studienwerk and Worcester College, Oxford.

%%%%%%%%%%%%%%%%%%%%%%%%%%%%%%%%%%%%%%%%%%%%%%%%%%%%%%%%%%
%
\begin{appendix}
\section{$\!\!\!\!\!\!\!$ppendix}
From $\nu$-DIS we obtain the following bounds, explicitly retaining
$\chi_{q_{i\!n}q_{o\!u\!t}},\eta_{exp},\widetilde{\chi}$:
\begin{eqnarray}
%|\lam^{\prime*}_{31k}~\lam^\prime_{21k}  |&\leq&\sqrt{
%\frac{32\times 1.77\times10^7}{2514178+6292146+91181+123871}  }~\times~
%\sqrt{\frac{\eta_{exp}}{{\widetilde{\chi}}}~~}~\Bigg(\frac{m_{\widetilde{
%{d_R}^k}}}{100~\mbox{GeV}}
%\Bigg)^2,  \nonumber\\ \nonumber\\
|\lam^{\prime*}_{31k}~\lam^\prime_{22k}  |&\leq&
\frac{1}{\sqrt{ 0.055~\chi_{su}+0.028~\chi_{u^Cs^C}}  }~\times~
\sqrt{\frac{\eta_{exp}}{{\widetilde{\chi}}}~~}~\Bigg(\frac{m_{\widetilde{
{d_R}^k}}}{100~\mbox{GeV}}
\Bigg)^2,  \nonumber\\\nonumber\\
|\lam^{\prime*}_{31k}~\lam^\prime_{23k}  |&\leq&~\frac{1}{\sqrt{
{0.00052~\chi_{bu}+0.0024~\chi_{u^Cb^C}}  }}~\times~\sqrt{\frac{\eta_{exp}}{
{\widetilde{\chi}}}}
~\Bigg(\frac{m_{\widetilde{{d_R}^k}}}{100~\mbox{GeV}}\Bigg)^2,  \nonumber
\\ \nonumber\\
%|\lam^{\prime*}_{32k}~\lam^\prime_{21k}  |&\leq&0.116639~\sqrt{\frac{
%32\times 1.77\times10^7}{2171646+5528383+13988+14031}  }~\times~\sqrt{
%\eta_{exp}/{\widetilde{\chi}}~~}~\Bigg(\frac{m_{\widetilde{{d_R}^k}}}
%{100~\mbox{GeV}}\Bigg)^2,
%  \nonumber\\ \nonumber\\
|\lam^{\prime*}_{32k}~\lam^\prime_{22k}  |&\leq&
\frac{1}{\sqrt{0.047~\chi_{sc}+0.0036~\chi_{c^Cs^C}}  }~\times~
\sqrt{\frac{\eta_{exp}}{{\widetilde{\chi}}}}~\Bigg(\frac{m_{\widetilde
{{d_R}^k}}}{100~\mbox{GeV}}
\Bigg)^2,  \nonumber\\ \nonumber\\
|\lam^{\prime*}_{32k}~\lam^\prime_{23k}  |&\leq&
\frac{1}{\sqrt{{0.00051~\chi_{bc}+0.00039~\chi_{c^Cb^C}}  }}~\times~\sqrt{
\frac{\eta_{exp}}{{\widetilde{\chi}}}}~\Bigg(\frac{m_{\widetilde{{d_R}^k}}}
{100~\mbox{GeV}}\Bigg)^2,\nonumber\\\nonumber\\
|\lam^{\prime*}_{k11}~\lam_{2k3}  |&\leq&\frac{1}{\sqrt{0.43~\chi_{du}
+0.036~\chi_{u^Cd^C}}  }~\times~\sqrt{\frac{\eta_{exp}}{{\widetilde{\chi}}}}
~\Bigg(\frac{m_{\widetilde{{\ell_L}^k}}}{100~\mbox{GeV}}\Bigg)^2,  \nonumber\\
 \nonumber\\
|\lam^{\prime*}_{k12}~\lam_{2k3}  |&\leq&\frac{1}{\sqrt{  0.023~\chi_{su}
+0.036~\chi_{u^Cs^C}}  }~\times~\sqrt{\frac{\eta_{exp}}{{\widetilde{\chi}}}}
~\Bigg(\frac{m_{\widetilde{{\ell_L}^k}}}{100~\mbox{GeV}}\Bigg)^2,  \nonumber\\
 \nonumber\\
|\lam^{\prime*}_{k13}~\lam_{2k3}  |&\leq&\frac{1}{\sqrt{0.00010~\chi_{bu}
+0.0020~\chi_{u^Cb^C} } }~\times~\sqrt{\frac{\eta_{exp}}{{\widetilde{\chi}}}}
\Bigg(\frac{m_{\widetilde{{\ell_L}^k}}}{100~\mbox{GeV}}\Bigg)^2,  
\nonumber\\ \nonumber\\
|\lam^{\prime*}_{k21}~\lam_{2k3}  |&\leq&\frac{1}{\sqrt{0.52~\chi_{dc}
+0.0068~\chi_{c^Cd^C}  }}~\times~\sqrt{
\frac{\eta_{exp}}{{\widetilde{\chi}}}}~\Bigg(\frac{m_{\widetilde{{\ell_L}^k}}}
{100~\mbox{GeV}}\Bigg)^2,  \nonumber\\ \nonumber\\
|\lam^{\prime*}_{k22}~\lam_{2k3}  |&\leq&\frac{1}{\sqrt{0.030~\chi_{sc}
+0.0068~\chi_{c^Cs^C}  }}~\times~\sqrt{
\frac{\eta_{exp}}{{\widetilde{\chi}}}}~\Bigg(\frac{m_{\widetilde{{\ell_L}^k}}}
{100~\mbox{GeV}}\Bigg)^2,  \nonumber\\ \nonumber\\
|\lam^{\prime*}_{k23}~\lam_{2k3}  |&\leq&\frac{1}{\sqrt{0.00011~\chi_{bc}
+0.00042~\chi_{c^Cb^C}}  }~\times~\sqrt{\frac{\eta_{exp}}{{\widetilde{\chi}}}}
~\Bigg(\frac{m_{\widetilde{{\ell_L}^k}}}{100~\mbox{GeV}}\Bigg)^2.  \nonumber
\end{eqnarray}
\end{appendix}
%
%%%%%%%%%%%%%%%%%%%%%%%%%%%%%%%%%%%%%%%%%%%%%%%%%%%%
%
%


\begin{thebibliography}{99}
%
\bibitem{w} S.Weinberg, \emph{Phys.Rev.} {\bf{D26}}
 287 (1982).
%
\bibitem{Hall:1983id}
L.Hall, M.Suzuki, \emph{Nucl.Phys.}  {\bf B231} 419 (1984).
%
\bibitem{d} 
H.Dreiner, \emph{hep-ph/9707435}.
%
\bibitem{Bhattacharyya:1996nj} G.Bhattacharyya,
\emph{Nucl.Phys.Proc.Suppl.} {\bf 52A} 83 (1997);
\emph{hep-ph/9608415}.
%
\bibitem{add} B.Allanach, A.Dedes, H.Dreiner,
 \emph{Phys.Rev.}
  {\bf{D60}} 075014 (1999); \emph{hep-ph/9906209}.
%
\bibitem{bgh} V.Barger, G.Giudice, T.Han, \emph{Phys.Rev.} {\bf D40} 2987 (1989). 
%
\bibitem{NOMAD}
NOMAD Collaboration, 
J.Altegoer et al., 
\emph{Nucl.Instrum. Methods} {\bf A404} 96 (1998).
%
\bibitem{CHORUS}
CHORUS Collaboration, 
E.Eskut et al.,
\emph{Nucl.Instrum. Methods} {\bf A401} 7 (1997).
% 
%\cite{Dedes:2001zi}
\bibitem{Dedes:2001zi}
A.Dedes, H.Dreiner, P.Richardson,
\emph{Phys.Rev.} {\bf{D65}} 015001 (2002); \emph{hep-ph/0106199}.
%
\bibitem{Pesen:zx}
CHORUS Collaboration, E.Pesen, \emph{Nucl.Phys.Proc.Suppl.}  {\bf 70} 219 (1999).
%
\bibitem{p2}
NOMAD Collaboration, P.Astier et al., \emph{Phys.Lett.}  {\bf B453} 169 (1999).
%
\bibitem{Gninenko:2001id}
S.Gninenko, M.Kirsanov, N.Krasnikov, V.Matveev, \emph{hep-ph/0106302}.
%
%
\bibitem{ag} K.Agashe, M.Graesser, \emph{Phys.Rev.} {\bf{D54}} 4445 (1996);
 \emph{hep-ph/9510439}.
%
\bibitem{dm} H.Dreiner, P.Morawitz, \emph{Nucl.Phys.} {\bf{B503}} 
55 (1997); \emph{hep-ph/9703279}.
%
%
\bibitem{models}
P.Bin\'etruy, S.Lavignac, P.Ramond,
\emph{Nucl.Phys.} {\bf{B477}} 353 (1996); \emph{hep-ph/9601243}.
%
\bibitem{cd}
A.Chamseddine, H.Dreiner,
\emph{Nucl.Phys.} {\bf{B458}} 65 (1996); \emph{hep-ph/9504337}.
%
\bibitem{as} 
NOMAD Collaboration,
P.Astier et al., \emph{Nucl.Phys.}
 {\bf{B611}} 3 (2001); 
\emph{hep-ex/0106102}.
%
\bibitem{es} 
CHORUS Collaboration,
E.Eskut et al., \emph{Phys.Lett.}
 {\bf{B497}} 8 (2001). 
%
\bibitem{roberto}
R.Petti, Seminar given at DESY, March 2002, unpublished. 
%
\bibitem{dpt} H.Dreiner, G.Polesello, M.Thormeier; to appear in 
\emph{Phys.Rev.} {\bf{D}}; \emph{hep-ph/0112228}.
%
\bibitem{pdg} Particle Data Group, D.Groom et al., \emph{Eur.Phys.J.} 
{\bf{C15}} 1 (2000).\\
We have taken into account the update from Dec.12, 2001.
%
\bibitem{spectrum} 
\makebox{G.Collazuol, A.Ferrari, A.Guglielmi, P.Sala, \emph{Nucl.Instrum.
 Methods} {\bf A449} 609 (2000).}
%
\bibitem{mrst} A.Martin, R.Roberts, W.Stirling, R.Thorne; 
\emph{hep-ph/0110215}.
%








\end{thebibliography}
\end{document}